\documentclass[conference]{IEEEtran}
\IEEEoverridecommandlockouts

\usepackage{cite}
\usepackage{amsmath,amssymb,amsfonts}
\usepackage{graphicx}
\usepackage{textcomp}
\usepackage{xcolor}
\usepackage{multirow}
\usepackage{bbding}

\usepackage{algpseudocode}
\usepackage{algorithm}
\usepackage{booktabs}
\usepackage{soul}

\def\BibTeX{{\rm B\kern-.05em{\sc i\kern-.025em b}\kern-.08em
    T\kern-.1667em\lower.7ex\hbox{E}\kern-.125emX}}
\begin{document}

\title{MSEMG: Surface Electromyography Denoising with a Mamba-based Efficient Network\\
}


\author{\IEEEauthorblockN{
Yu-Tung Liu\IEEEauthorrefmark{1}\IEEEauthorrefmark{4},
Kuan-Chen Wang\IEEEauthorrefmark{2}\IEEEauthorrefmark{4},
Rong Chao\IEEEauthorrefmark{4}\IEEEauthorrefmark{5}, Sabato Marco Siniscalchi\IEEEauthorrefmark{3},
Ping-Cheng Yeh\IEEEauthorrefmark{2},
and Yu Tsao\IEEEauthorrefmark{4}}
\IEEEauthorblockA{
\IEEEauthorrefmark{1}Department of Electronics and Electrical Engineering, National Yang Ming Chiao Tung University, Taiwan\\
\IEEEauthorrefmark{2}Graduate Institute of Communication Engineering, National Taiwan University, Taiwan \\
\IEEEauthorrefmark{5}Department of Computer Science and Information Engineering, National Taiwan University, Taiwan \\
\IEEEauthorrefmark{3}University of Palermo, Italy
\IEEEauthorrefmark{4}Research Center for Information Technology Innovation, Academic Sinica, Taiwan\\
Email: tonyliu.ee09@nycu.edu.tw,\ 
d12942016@ntu.edu.tw,\
roychao@cmlab.csie.ntu.edu.tw\\
sabatomarco.siniscalchi@unipa.it,\
pcyeh@ntu.edu.tw,\
yu.tsao@citi.sinica.edu.tw
}
}

\maketitle
\begin{abstract}
Surface electromyography (sEMG) recordings can be contaminated by electrocardiogram (ECG) signals when the monitored muscle is closed to the heart. Traditional signal processing-based approaches, such as high-pass filtering and template subtraction, have been used to remove ECG interference but are often limited in their effectiveness. Recently, neural network-based methods have shown greater promise for sEMG denoising, but they still struggle to balance both efficiency and effectiveness. In this study, we introduce MSEMG, a novel system that integrates the Mamba state space model with a convolutional neural network to serve as a lightweight sEMG denoising model. We evaluated MSEMG using sEMG data from the Non-Invasive Adaptive Prosthetics database and ECG signals from the MIT-BIH Normal Sinus Rhythm Database. The results show that MSEMG outperforms existing methods, generating higher-quality sEMG signals using fewer parameters. 
\end{abstract}


\section{Introduction}

\label{sec:intro}
Surface electromyography (sEMG) captures electrical motor nerve signals and converts them into digital recordings. sEMG signals are measured noninvasively using electrodes and provide rich clinical insights into muscular activities. sEMG has been applied in a variety of applications, including neuromuscular system investigation~\cite{tang2018novel}, rehabilitation~\cite{engdahl2015surveying}, stress monitoring~\cite{wijsman2013wearable}, the assessment of neuromuscular and respiratory disorders~\cite{domnik2020clinical, vandenbussche2015assessment}, prosthesis control~\cite{ma2014hand}, and gesture recognition in virtual reality (VR)~\cite{cote2021transferable,xiao2023interactive}. However, sEMG signals can be contaminated by electrocardiography (ECG) signals when sensors are placed near the thorax.  This interference reduces the quality and fidelity of sEMG recordings, which can affect clinical outcomes and human-computer interactions.


sEMG and ECG signals share a frequency band of 0 to 100 Hz~\cite{winter2009biomechanics}, making it difficult to eliminate ECG artifacts from sEMG. Some conventional signal processing methods, such as high pass filters (HP) and template subtraction (TS)~\cite{xu2020comparative,drake2006elimination}, struggle to extract clean sEMG signals, particularly when dealing with low signal-to-noise ratios (SNR). Given these challenges, neural networks (NN) have been introduced into sEMG enhancement techniques owing to their powerful nonlinear mapping capabilities and data-driven characteristics~\cite{wang2023ecg, liu2024sdemg, wang2024trustemg}. For example, ~\cite{wang2023ecg} proposed a fully convolutional network (FCN), a discriminative model that outperformed the traditional methods (i.e., HP and TS) for the removal of ECG artifacts, and ~\cite{liu2024sdemg} introduced a score-based diffusion model, a generative approach that achieved superior results,  compared with those achieved using previously developed techniques. However, diffusion models incur high computational costs during inference, which limits their use in real-time or resource-constrained sEMG applications, thus necessitating an efficient and high-performance sEMG denoising method.

The Mamba state space model (SSM)~\cite{gu2023mamba} is a prominent recurrent neural network (RNN) that was designed to effectively capture temporal information from long sequences with linear time complexity. Its low computational cost distinguishes Mamba from Transformer~\cite{vaswani2017attention}, which also excels in handling global dependencies, but requires more expensive quadratic time complexity. Owning to its efficiency, Mamba has been applied in various domains, including large language models~\cite{lieber2024jamba} and signal enhancement~\cite{chao2024investigation}. However, its potential for sEMG denoising has not yet been explored, as previous methods have primarily relied on convolutional neural networks (CNNs).

In this study, we introduced MSEMG, which is a lightweight Mamba-inherited CNN designed to efficiently generate high-quality sEMG signals. By integrating Mamba with CNNs, MSEMG overcomes the limited receptive fields of the convolutional layers and can capture both local and long-range dependencies. Experimental results demonstrate that MSEMG outperforms previous sEMG denoising techniques on several evaluation metrics. Furthermore, MSEMG adopts fewer parameters than existing state-of-the-art sEMG denoising methods. These findings suggest that MSEMG is a promising solution for providing high-quality signals for various sEMG applications. To the best of our knowledge, this is the first study to explore the application of Mamba to sEMG processing.

The remainder of this paper is organized as follows. Section 2 reviews related works. Section 3 introduces the proposed approach. Section 4 presents the experimental setup and results. Finally, Section 5 concludes the paper and discusses future work.

\section{Related Work}
\label{sec:related}

\subsection{ECG interference removal methods}
\label{ssec:previous}

Traditional methods, such as HP and TS\cite{xu2020comparative,drake2006elimination}, use classical signal processing techniques to remove ECG artifacts from sEMG signals. However, these methods exhibit limitations in achieving high-quality sEMG recordings. HP removes the frequency band associated with ECG interference and removes low-frequency components of the sEMG. TS, on the other hand, operates in the time domain to handle ECG interference by extracting ECG templates by filtering or waveform averaging\cite{xu2020comparative,junior2019template}. However, TS may be less effective in real-world applications, as it assumes that the sEMG signals follow a zero-mean Gaussian distribution. This study implements both HP and TS methods for comparison in our experiments.

NN-based techniques have demonstrated significant improvement in sEMG denoising. Wang et al.~\cite{wang2023ecg} introduced an FCN as a denoising autoencoder to remove ECG artifacts from sEMG signals. The FCN consists of two main components: an encoder to extract features and a decoder to translate the latent representation back into the signal. This approach has yielded impressive results in various research areas, and FCNs have proven to be more effective than traditional methods. However, FCN-based methods still struggle with signal distortion. 

To address these issues, SDEMG was introduced~\cite{liu2024sdemg}, applying the generative model to the problem of sEMG denoising. Diffusion models, known for generating high-quality and high-fidelity outputs, were used to identify noise during the diffusion process and remove ECG interference during sampling. Although SDEMG is highly effective, repeated inference steps lead to high computational costs, rendering them impractical for use in resource-limited sEMG applications. 

\subsection{Mamba State Space Model}
\label{ssec: mamba}

The Mamba State Space Model is a cutting-edge NN architecture designed for sequence modeling, particularly for managing long sequences~\cite{gu2023mamba}. Mamba addresses some of the limitations of traditional Transformer models, particularly in terms of computational efficiency and performance in various data modalities, such as language~\cite{lieber2024jamba}, audio~\cite{chao2024investigation,shams2024ssamba}, and genomics~\cite{zhang2024chimamba, thoutammsamamba}. Although state space models previously exhibited limited performance, via the selective state space mechanism, Mamba can perform content-based reasoning, enabling Mamba to selectively propagate or neglect information based on the current token. Moreover, Mamba offers linear scaling in sequence length and achieves faster inference, exhibiting a throughput up to five times higher than that of Transformers. As a result, Mamba has demonstrated state-of-the-art performance across several modalities, outperforming Transformers of the same size and matching those twice their size in both pre-training and downstream evaluation. Initially, the Mamba-3B model outperformed comparable Transformer models in language tasks.

\section{The Proposed Method}
\label{sec:proposed}

In this section, we introduce the Mamba model and the selection mechanism as a key proportion of this novel integration of CNN and the state space model. The implementation detail of the NN in this work will be further elaborated. 

\begin{figure}[t!]
    \centering
    \includegraphics[width=\linewidth]{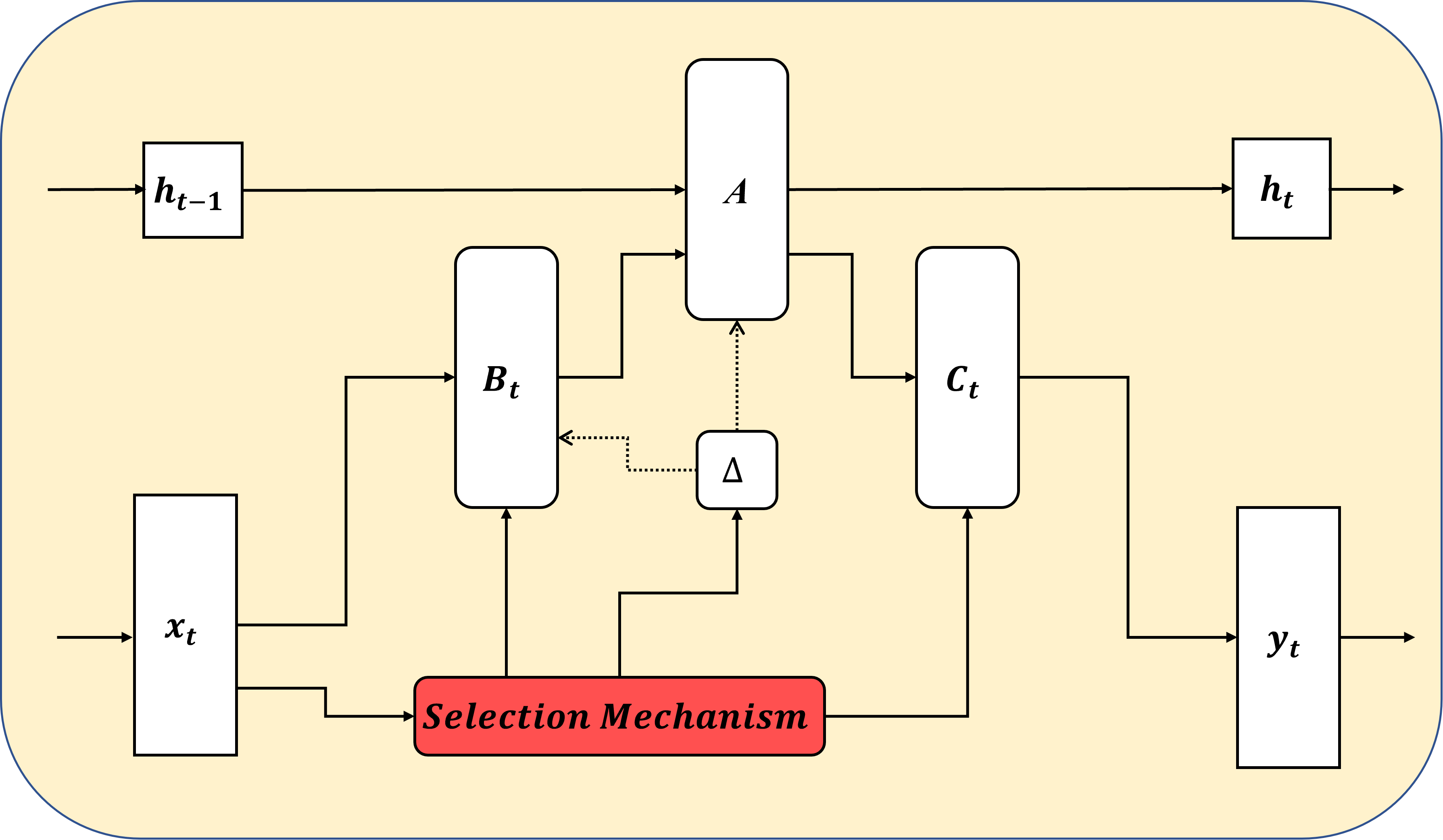}
    \caption{The Selective State Space Model.}
    \label{fig: selection}
\end{figure}

\begin{figure*}[h!]
    \centering
    \includegraphics[width=.9\textwidth]{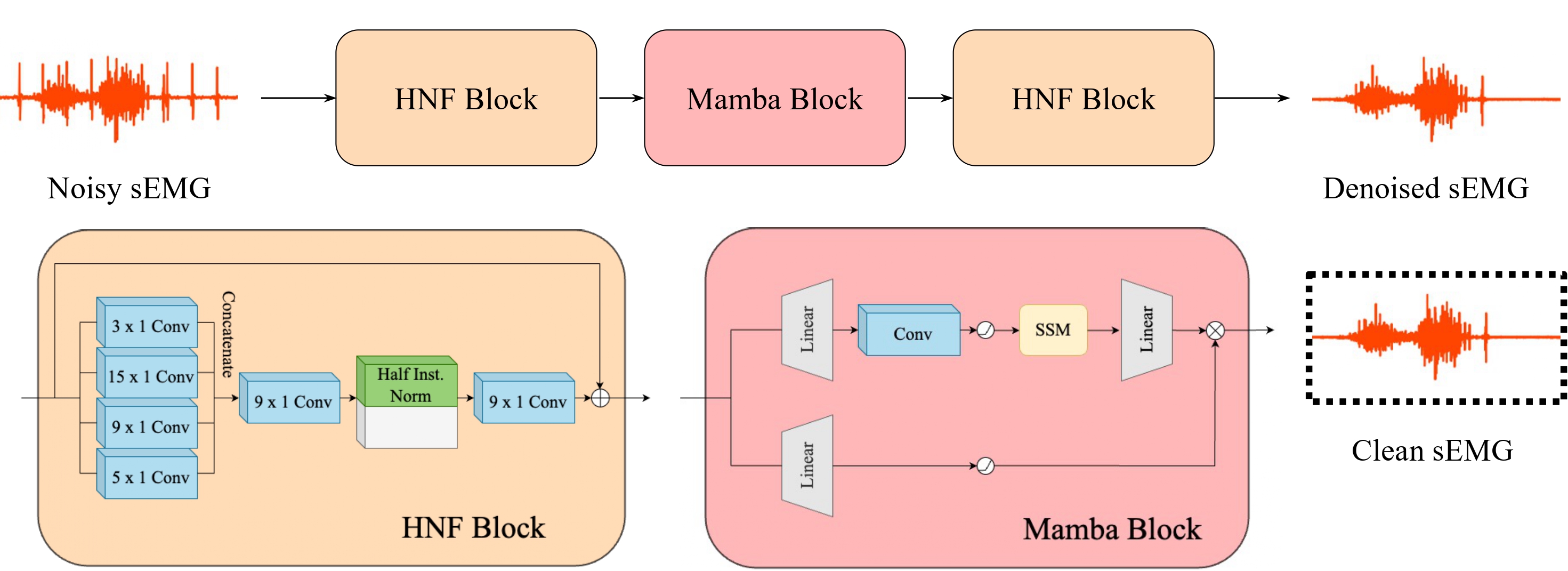}
    \caption{The denoising framework and the model architecture of MSEMG.}
    \label{fig: model_arch}
\end{figure*}

\subsection{Selective state space model}
\label{ssec: ssm}
The Mamba SSM (Fig. \ref{fig: selection}) introduced a selective scan mechanism to perform adaptive sequence mapping from input $x(t) \in \mathbb{R}$ to $y(t) \in \mathbb{R}$. There are some key elements in the SSM: a trainable parameter $\Delta$, a state transition matrix $A \in \mathbb{R}^{H \times H}$, an input projection matrix $B \in \mathbb{R}^{H \times 1}$, an output projection matrix $C \in \mathbb{R}^{1 \times H}$, and a hidden state $h \in R^{H}$. Note that $H$ indicates the dimension of the hidden state. Empirically, the dimension of the hidden state is determined by the complexity of the problem. The state transition process could be described by Eq. (\ref{eq: state continuous}) and the projection of the output by Eq. (\ref{eq: output continuous}).

\begin{equation}
h^\prime(t) = Ah(t) + Bx(t)
\label{eq: state continuous} 
\end{equation}

\begin{equation}
y(t) = Ch(t)
\label{eq: output continuous} 
\end{equation}

The above equations describe state transitions in continuous-time signals. However, discrete-time signals dominate actual experiments and downstream applications. The terms $\overline{A}$ and $\overline{B}$ in Eqs. (\ref{eq: state discrete}) and (\ref{eq: output discrete}) are the discretized matrices approximated by a zero-order hold. 

\begin{equation}
h_{t} = \overline{A}h_{t-1} + \overline{B}x_{t}
\label{eq: state discrete} 
\end{equation}

\begin{equation}
y_{t} = Ch_{t}
\label{eq: output discrete} 
\end{equation}

The learnable parameter $\Delta$ adjusts the balance between the hidden state and the current input:

\begin{equation}
\overline{A} = exp(\Delta A)
\label{eq: discretization A} 
\end{equation}

\begin{equation}
\overline{B} = (\Delta A)^{-1}(exp(\Delta A) - I) \cdot \Delta B
\label{eq: discretization B} 
\end{equation}

Recursion can be observed in Eqs. (\ref{eq: state discrete} and \ref{eq: output discrete}). Therefore, a convolution kernel $\overline{K}$ is introduced to reduce the computational workload. 

\begin{equation}
\overline{K} = (C\overline{B},C\overline{AB}, ...,C\overline{A}^{k}\overline{B}) 
\label{eq: unroll} 
\end{equation}

\begin{equation}
y = x * \overline{K}
\label{eq: unroll out} 
\end{equation}

One of the core concepts of the Mamba model is input awareness; that is, $\overline{K}$ is constantly updated by the parameter $\Delta$. Hence, we could only set $k=0$ in Mamba. Subsequently, $\overline{A}$, $\overline{B}$, and $C$ are also changing. The selective scan algorithm proposed by Gu and Dao~\cite{gu2023mamba} efficiently accommodates these changes.

\subsection{Model architecture}
\label{subsec: model}

Fig. \ref{fig: model_arch} shows the denoising framework of MSEMG. The key idea behind MSEMG is to leverage the unique capabilities of the Mamba block to generate high-quality output. The process begins with input sEMG signals projected into the latent space through the Half Normalized Filters (HNF) block~\cite{romero2021deepfilter} that applies various convolutional kernels to extract features at different resolutions. These features are concatenated and further processed via channel-wise convolution to match the wide frequency range of sEMG signals. Next, the Mamba block models and processes the sequence within the latent space as described in the previous subsection. Finally, the second HNF block reconstructs the denoised sEMG signal, producing the final output. The figure illustrates the denoising process by showing noisy, denoised, and clean sEMG. The noisy waveform represents sEMG signals under ECG interference at an SNR of -4 dB.

\section{Experiments}
\subsection{Dataset preprocessing and preparation}
The sEMG signals used in this study were sourced from the DB2 subset of the Non-Invasive Adaptive Prosthetics (NINAPro) database~\cite{atzori2014electromyography}, which consists of 12-channel sEMG recordings of hand movements from 40 subjects. These recordings were obtained from the upper limbs. The DB2 subset includes three sessions, namely, Exercises 1, 2, and 3, which feature 17, 22, and 10 movements, respectively. Each movement is repeated six times for five seconds, followed by a three-second rest interval. To remove potential noise in the sEMG data, a fourth-order Butterworth bandpass filter with cutoff frequencies of 20 and 500 Hz~\cite{machado2021deep,liu2024sdemg} was applied. The signals were then downsampled to 1 kHz, normalized, and segmented into 10 s intervals.

For ECG interference, we used the MIT-BIH Normal Sinus Rhythm Database (NSRD) from PhysioNet~\cite{goldberger2000physiobank}, which contains 2-channel ECG recordings from 18 healthy subjects sampled at 128 Hz. Channel 1 ECG data were filtered using third-order Butterworth high-pass and low-pass filters with cutoff frequencies of 10 and 200 Hz to eliminate potential noise. Previous studies have also used this dataset to simulate ECG interference in sEMG signals~\cite{machado2021deep,wang2023ecg,liu2024sdemg}.

Following previous work~\cite{liu2024sdemg}, sEMG segments of Channel 2, Exercise 1 and 3, involving 30 subjects, were used for training and validation. Each training segment used 10 randomly selected ECG signals from 12 subjects in the MIT-BIH NSRD, superimposed as artifacts at six different SNR levels (-15 to -5 dB in 2 dB increments). The validation set used ECG artifacts from three other subjects with the same SNR levels. For testing, we evaluated the generalizability of the model using different subjects, movements, channels, and SNR levels. Test data were obtained from Channels 9, 10, 11, and 12, Exercise 2, involving 10 subjects. ECG signals from three remaining subjects (i.e., 16420, 16539, and 16786) were used as interference, with SNR levels ranging from -14 to 0 dB in 2 dB increments. 

\subsection{Evaluation metrics}
\label{ssec:metrics}
Performance was evaluated based on signal reconstruction quality and feature extraction accuracy~\cite{xu2020comparative,chiang2019noise,wang2023ecg,liu2024sdemg}. The signal reconstruction quality was measured by SNR improvement (SNR${imp}$) and root-mean-square error (RMSE), which calculate the difference between input and output SNRs and the error between output signals and ground truth, respectively. In addition, the RMSE of the average rectified value (ARV) and mean frequency (MF) feature vectors were used to assess sEMG feature extraction accuracy~\cite{xu2020comparative,wang2023ecg}. These metrics were calculated following the methods in previous studies~\cite{xu2020comparative,wang2023ecg,liu2024sdemg}. Higher SNR${imp}$ values, lower RMSE values, and lower RMSE values for the ARV and MF indicate better signal reconstruction and higher fidelity.

\subsection{Results and discussion}
\label{ssec:result}

\begin{figure}[t!]
    \centering
    \includegraphics[width=\linewidth]{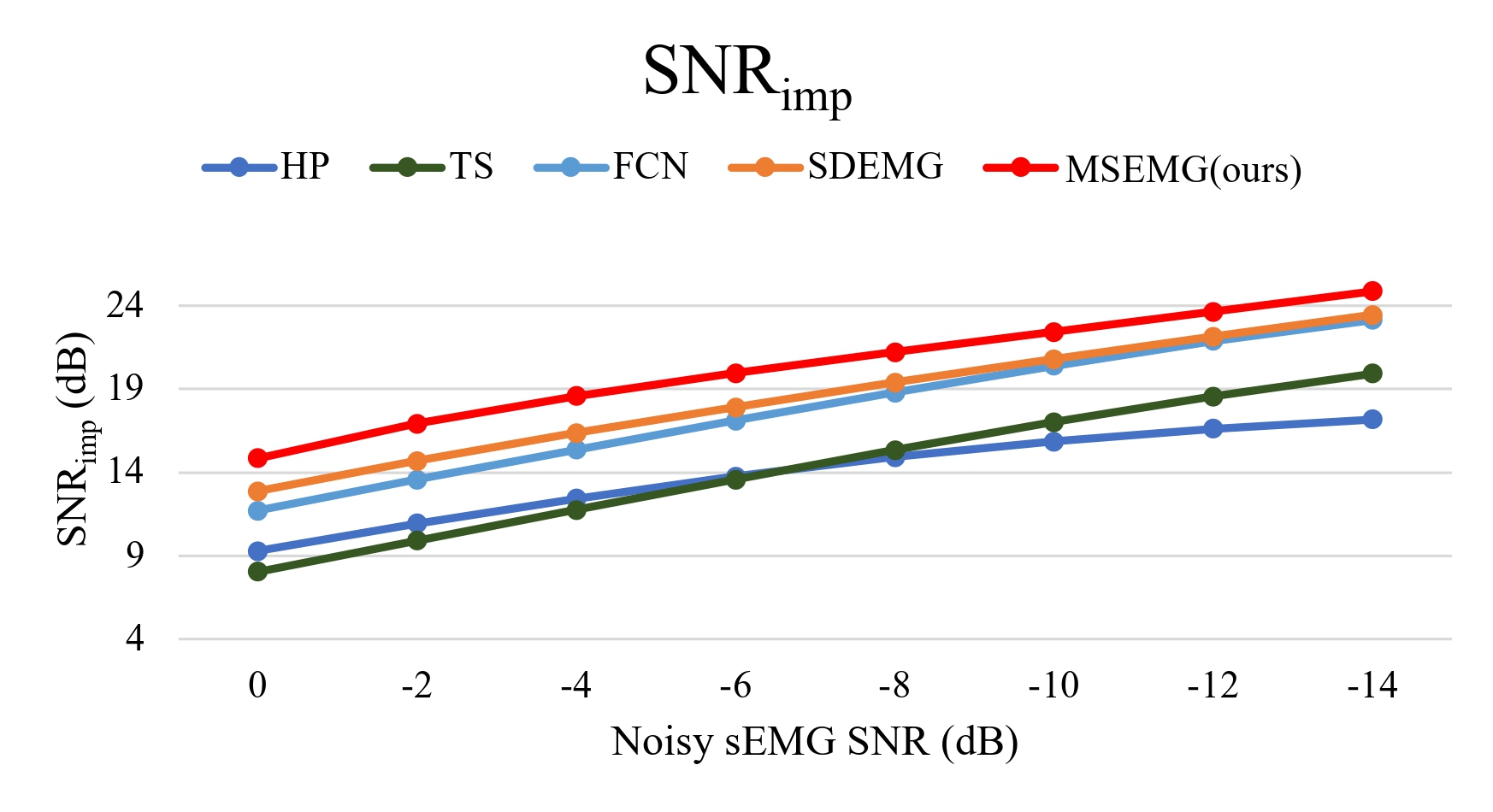}
    \caption{Comparison of denoising methods with SNR$_{imp}$ under various SNR conditions.}
    \label{fig: big}
\end{figure}

We compared the performance of MSEMG with two traditional methods, namely HP and TS, as well as two NN-based approaches, namely FCN~\cite{wang2023ecg} and SDEMG~\cite{liu2024sdemg}. Table~\ref{tab: result-table} summarizes the overall results for the various evaluation metrics. MSEMG consistently outperforms all other methods, achieving the highest SNR${imp}$, the lowest RMSE, and the lowest RMSE values for ARV and MF features. Fig.~\ref{fig: big} further illustrates the SNR${imp}$ performance under different SNR conditions, highlighting that MSEMG produces superior output quality across all levels of noise interference. These results suggest that MSEMG offers a more robust and effective approach to sEMG denoising, largely because of its enhanced ability to capture both local and global dependencies in the data.

\begin{table}[t!]
\centering
\small
\renewcommand\arraystretch{1.5}
\caption{Overall performance of HP, TS, FCN, SDEMG, and MSEMG.}
\smallskip
\label{tab: result-table}
\resizebox{1\linewidth}{!}{
\begin{tabular}{@{}clclc@{}}
\toprule
            & \multicolumn{1}{c}{SNR$_{imp}$}(dB)& RMSE        & \multicolumn{1}{c}{RMSE$_{ARV}$} & RMSE$_{MF}$ (Hz)   \\ \midrule
HP          & 13.885                   & 1.735e-2              & 3.064e-3             & 19.471               \\
TS          & 14.279                   & 1.626e-2              & 3.859e-3             & 23.149               \\
FCN         & 17.758                   & 1.178e-2              & 3.864e-3             & 18.038               \\ 
SDEMG       & 18.467                   & 1.138e-2              & 2.809e-3             & 14.435               \\
\textbf{MSEMG(Ours)} & \textbf{20.317} & \textbf{8.603e-3}   & \textbf{2.382e-3}    & \textbf{11.379}       \\ \midrule
\end{tabular}
}
\end{table}

\begin{figure}[t]
    \centering
    \includegraphics[width=.9\linewidth]{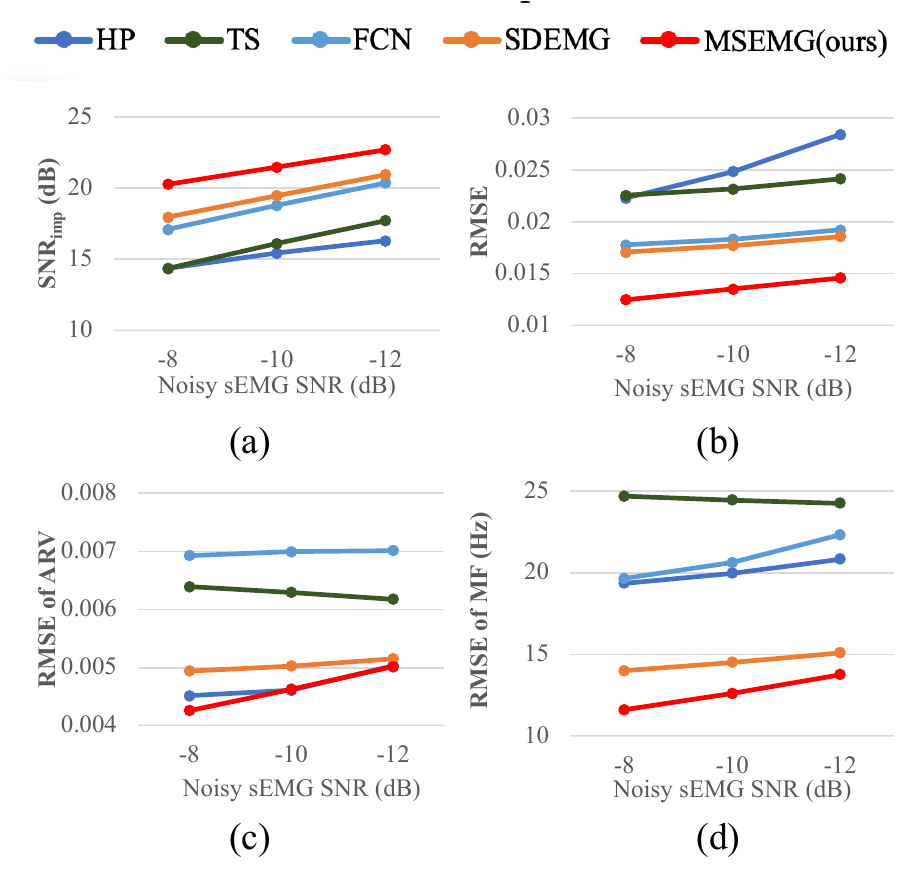}
    \caption{Performance under the scenario simulating trunk sEMG with ECG contamination.}
    \label{fig: small}
\end{figure}

Fig.\ref{fig: small} demonstrates the denoising performance in a specific scenario that simulates trunk sEMG with ECG contamination. In this case, we used the biceps brachii sEMG data (Channel 11 in NINAPro DB2) as simulation data, with the SNR set to approximately -10 dB, mimicking typical contamination conditions in the trunk sEMG\cite{xu2020comparative,drake2006elimination,zhou2006eliminating}. Under these conditions, MSEMG maintained its superiority in all evaluation metrics.

Table~\ref{tab: param} compares the number of trainable parameters for the NN-based denoising methods. MSEMG is approximately one-fifth the size of SDEMG, which was the previous state-of-the-art method. Given the superior performance of MSEMG in removing ECG artifacts and its smaller model size, we conclude that MSEMG offers an efficient and effective solution for sEMG denoising.

\begin{table}[t]
\centering
\caption{Comparison of model size.}
\smallskip
\label{tab: param}
\resizebox{\linewidth}{!}{
\begin{tabular}{@{}ccccc@{}}
\toprule

& FCN \cite{wang2023ecg} & SDEMG \cite{liu2024sdemg} & \textbf{MSEMG(Ours)} \\ \midrule
\# of Parameters & 137,801 & 1,233,857 & 279,937 \\ \midrule
\end{tabular}
}
\end{table}

\section{Conclusion}
\label{sec:conclusion}
In this study, we present MSEMG, which is a novel denoising model that combines the Mamba state space model with CNNs. MSEMG effectively captures local and global dependencies in sEMG signals while maintaining low computational costs. Our experimental results demonstrate that MSEMG consistently outperforms previous methods across all evaluation metrics, even under varying SNR conditions. These findings highlight MSEMG as a promising solution for sEMG denoising for a wide range of applications. Future work will focus on further optimizing MSEMG performance with additional data and more complex denoising scenarios. We also plan to apply MSEMG to downstream tasks such as hand gesture recognition and respiratory estimation to assess its effectiveness in real-world applications.
\vfill\pagebreak

\bibliographystyle{IEEEbib}
{\small 
\bibliography{refs}}

\end{document}